\documentclass{article}

\usepackage[final]{neurips_2025_ml4ps}
\usepackage{graphicx}
\usepackage{subcaption} 
\usepackage[utf8]{inputenc} 
\usepackage[T1]{fontenc}    
\usepackage{hyperref}       
\usepackage{url}            
\usepackage{booktabs}       
\usepackage{amsfonts}       
\usepackage{nicefrac}       
\usepackage{microtype}      
\usepackage{xcolor}         
\usepackage{graphicx}
\usepackage{amsmath}
 
\usepackage{siunitx}
\usepackage[table]{xcolor} 
\usepackage{tabularx}
\usepackage{threeparttable}

\definecolor{headgray}{gray}{0.92}
\definecolor{rowgray}{gray}{0.97}

\usepackage[percent]{overpic} 
\usepackage[font=small,labelfont=bf]{caption}

\usepackage{graphicx}
\usepackage{subcaption}
\usepackage{subcaption}
\usepackage[font=small,labelfont=bf]{caption}

\title{Formatting Instructions For NeurIPS 2025}

\title{From Images to Physics: Probabilistic Inference of Galaxy Parameters and Emission Lines via VAE–Normalizing Flows}

\author{%
  Adiba Amira Siddiqa \\
  Department of Physics\\
  Bryn Mawr College\\
  Bryn Mawr, PA 19010 \\
  \texttt{asiddiqa@brynmawr.edu} \\
  \And
  Sayed Shafaat Mahmud \\
  Department of Physics and Astronomy\\
  Colgate University\\
  Hamilton, NY 13346 \\
  \texttt{smahmud@colgate.edu} \\
  \And
  Rafael Martinez-Galarza
  \\
  AstroAI\\
  Harvard-Smithsonian Center for Astrophysics\\
  Cambridge, MA 02138\\
  \texttt{jmartine@cfa.harvard.edu} \\
}

\begin{document}

\maketitle

\begin{abstract}
We introduce a Variational Autoencoder (VAE)--Normalizing Flow (NF) framework for rapid probabilistic inference of galaxy properties and emission line fluxes at $z \leq 0.3$ from SDSS \textit{gri} imaging and photometry. Our model probabilistically infers stellar mass, star formation rate (SFR), redshift, gas-phase metallicity, and central black hole mass for a given galaxy. The model accruacy matches current non-spectroscopic methods for stellar mass and redshift, surpasses them for SFR and metallicity, and introduces the first probabilistic central black hole mass estimates from imaging + photometry. It also delivers probabilistic estimates of H$\alpha$, H$\beta$, [N~\textsc{ii}], and [O~\textsc{iii}] emission line fluxes directly from imaging, enabling SFR, metallicity, dust, and AGN/shock diagnostics without spectroscopy. This approach opens new pathways for scalable, physics-informed inference in upcoming surveys such as Roman and Rubin LSST.
\end{abstract}

\section{Introduction}

Inferring the physical properties of galaxies such as star formation rate (SFR) \citep{weidner2004implications, martin2005star}, stellar mass \citep{weidner2004implications, weaver2023cosmos2020}, redshift \citep{masters2015mapping, gray2008review}, gas-phase metallicity \citep{carton2018first, pasquali2012gas}, and central black hole mass \citep{beifiori2012correlations, gebhardt2000relationship} is fundamental to understanding galaxy formation and evolution \citep{mo2010galaxy, cattaneo2009role, buonomo2000galaxy, pasquali2010ages}. These parameters govern how galaxies assemble their stellar mass, regulate star formation, and evolve across different environments. In addition, emission line fluxes such as H$\alpha$, H$\beta$, [N~\textsc{ii}], and [O~\textsc{iii}] are central to constraining SFR \citep{gallagher1989star, tacchella2022h}, metallicity \citep{groves2006emission, corlies2016empirically}, dust content \citep{hu1992ly}, and form the basis of key diagnostics such as the BPT diagram \citep{baldwin1981classification, kewley2013cosmic}. However, measuring these emission lines requires spectroscopy, which is observationally expensive and difficult for the billions of galaxies expected in upcoming large surveys.

Traditional approaches for parameter inference rely on spectral energy distribution (SED) fitting pipelines such as \textsc{Prospector} \citep{johnson2021stellar}, \textsc{Bagpipes} \citep{carnall2018inferring}, or \textsc{CIGALE} \citep{boquien2019cigale}. While physically grounded, these methods are computationally intensive. In response, recent studies have explored deep learning methods that directly regress galaxy properties from photometry and imaging \citep{hoyle2016measuring, zeraatgari2024exploring, surana2020predicting, li2022galaxy} . \textsc{AstroCLIP} \citep{parker2024astroclip}, for example, introduced a multimodal contrastive learning framework that aligns imaging and photometry in a shared latent space, enabling flexible downstream predictions across surveys. \cite{gagliano2023physics} introduced a conditional VAE for rapid inference of stellar mass and redshift. Yet, despite these advances, most existing approaches output only point estimates, lack calibrated uncertainty, and rarely attempt joint inference of both physical properties and emission line fluxes from imaging alone.

Our work addresses these gaps. We introduce a VAE--Normalizing Flow framework for rapid probabilistic inference of galaxy properties and emission line fluxes at $z \leq 0.3$ from SDSS \textit{gri} imaging and photometry. The first flow infers posteriors for stellar mass, SFR, redshift, gas-phase metallicity, and central black hole mass. The model matches or exceeds current non-spectroscopic methods for stellar mass, redshift, and metallicity, surpasses them for SFR, and introduces the first probabilistic central black hole mass estimates from imaging plus photometry. The second flow predicts posterior distributions for H$\alpha$, H$\beta$, [N~\textsc{ii}], and [O~\textsc{iii}] fluxes directly from imaging and photometry, enabling SFR, metallicity, dust, and AGN/shock diagnostics without spectroscopy. Even when using only photometry, our NF achieves accuracy comparable to or exceeding that reported in the literature for these parameters. This joint framework produces well-calibrated posteriors, operates over 100$\times$ faster than SED fitting, and opens new pathways for scalable, physics-informed inference in upcoming surveys such as the Roman Space Telescope \citep{mosby2020properties} and Rubin LSST \citep{hambleton2023rubin}.

\section{Methodology}

\begin{figure}[htbp]
  \centering
  \includegraphics[width=0.70\textwidth]{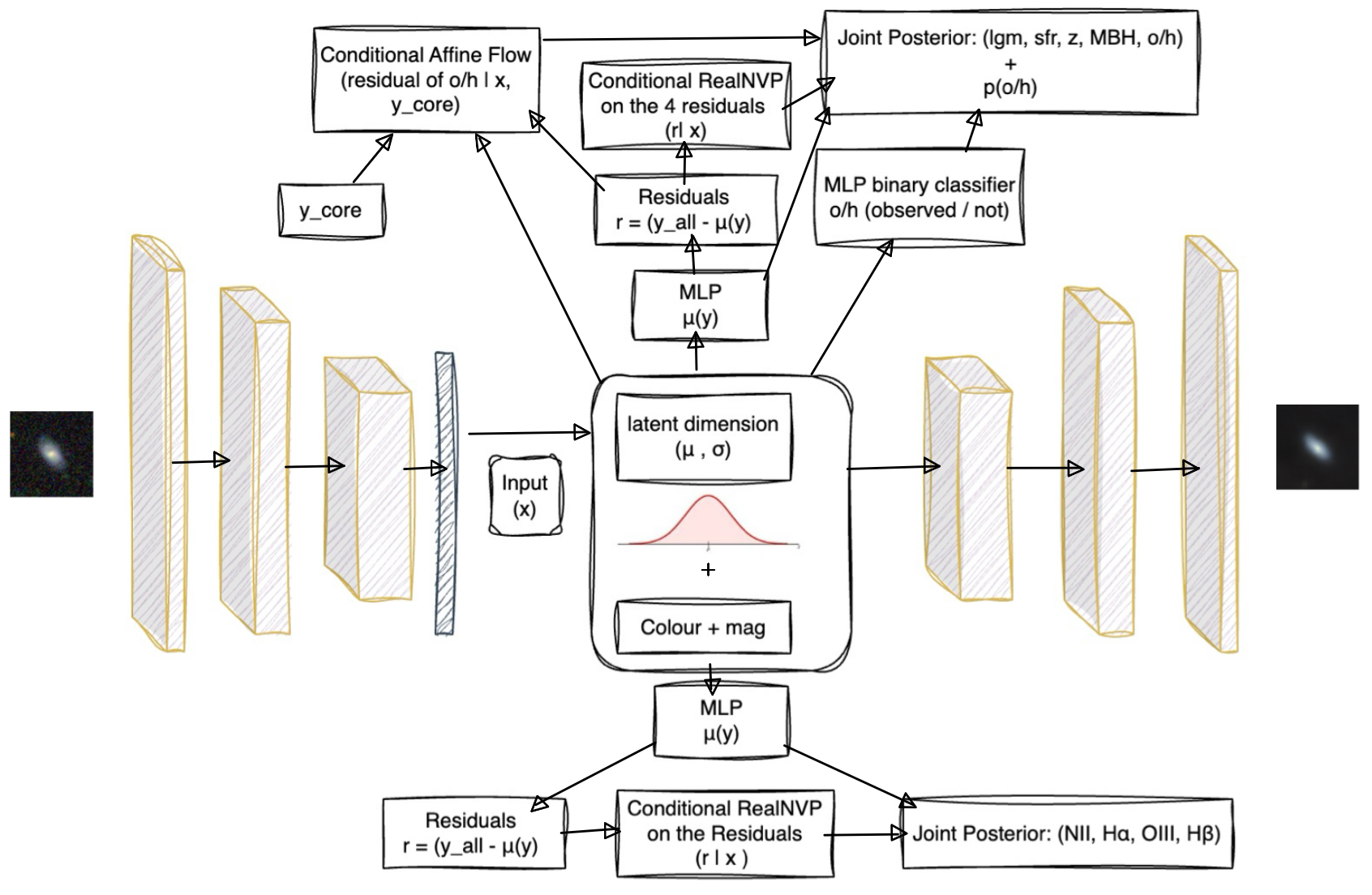}
  \caption{\textbf{VAE–NF hybrid architecture}}
  \label{fig:vae-dnn-architecture}
\end{figure}

\subsection{Data Selection and Preprocessing}

We select approximately 250{,}000 galaxies from the SDSS Main Galaxy Sample \citep{strauss2002spectroscopic} with spectroscopically measured redshifts $z \leq 0.3$ and available photometry in the $u,g,r,i,z$ bands. For each galaxy, we construct 3-channel $160 \times 160$ \textit{gri}-band cutout images centered on the galaxy using archival SDSS imaging. We also compute four rest-frame color indices: $u-g$, $g-r$, $r-i$, and $i-z$, which encode integrated photometric information. 

Of the full sample, roughly 100{,}000 galaxies are used to train and validate the VAE, while the remaining $\sim$150{,}000 are reserved for the NF stage. After discarding sources with missing observations, we are left with approximately 125{,}000 galaxies for NF training, validation, and testing on a 70/15/15 split.

The target parameters for the first NF are stellar mass, star formation rate (SFR), redshift, gas-phase metallicity, and central black hole mass, with the latter inferred from the stellar velocity dispersion using the empirical $M_{\mathrm{BH}}$–$\sigma$ relation of \cite{mcconnell2013revisiting}. The second normalizing flow predicts posterior distributions for the emission line fluxes H$\alpha$, H$\beta$, [N~\textsc{ii}]~(6583), and [O~\textsc{iii}]~(5007). All physical parameters and emission line measurements are drawn from SDSS spectroscopic and catalog-based measurements, standardized to zero mean and unit variance for training stability. 

\subsection{Variational Autoencoder Architecture}
We train the VAE \citep{kingma2019introduction} to learn a 32-dimensional latent representation of each galaxy’s imaging data.. The encoder has three convolutional layers (kernel size 4, stride 2, padding 1) followed by fully connected layers producing the latent mean $\mu \in \mathbb{R}^{32}$ and log-variance $\log \sigma^2 \in \mathbb{R}^{32}$. Latent vectors $\mathbf{z}$ are sampled via the reparameterization trick $\mathbf{z} \sim \mathcal{N}(\mu, \sigma^2)$. The decoder rebuilds the input images from the latent representation through a series of transposed convolutional layers

The VAE is optimized with mean squared reconstruction loss plus KL divergence to a standard Gaussian prior, using Adam with learning rate $10^{-4}$. After convergence, the encoder is frozen to extract latent features for all galaxy images.

\subsection{Two-Stage Conditional Normalizing Flow Framework}

We implement a two-stage probabilistic framework built on top of a vanilla VAE trained on galaxy images. The inputs to our model combine the 32 latent means and 32 standard deviations extracted from the VAE with the observed photometric colors and apparent magnitudes. A plain MLP encoder first processes these inputs into a 256-dimensional latent representation.From this representation, we construct two downstream branches. The first branch is an MLP that predicts the mean estimates for five physical parameters: stellar mass ($M_\star$), star formation rate (SFR), redshift ($z$), central black hole mass ($M_{\mathrm{BH}}$), and gas-phase metallicity, and compute their residuals. For the four core parameters ($M_\star$, SFR, $z$, and $M_{\mathrm{BH}}$) we train a conditional RealNVP\citep{dinh2016density} flow with 12 affine coupling layers to model the distrbution of the residuals. Each coupling layer uses a small MLP to compute the scale and translation parameters (\(s\), \(t\)) allowing the flow to learn a complex, invertible transformation from the residual space to a standard Gaussian. This conditional flow learns the joint probability distribution of the residuals given the encoded representation from the MLP.

The second branch focuses on the residuals of metallicity using a one-dimensional conditional affine flow. The flow is conditioned on the encoded representation and the true core parameters during training, allowing it to learn how metallicity depends on the other physical quantities. During inference, it conditions on the sampled core parameters (computed as the mean estimates plus the sampled residuals) from the first branch instead of the true values, thereby implementing the chain-rule factorization:
\[
p(y_{\mathrm{core}}, \mathrm{O/H} \mid x) = p(y_{\mathrm{core}} \mid x) \, p(\mathrm{O/H} \mid y_{\mathrm{core}}, x).
\]
We also use a small MLP with a sigmoid output to predict whether a galaxy has a measurable metallicity; it takes only the shared encoded representation as input and is trained jointly with the first branch using a combined loss (MSE for regression and binary cross-entropy for metallicity detectability).

Sampling from both flows produces calibrated joint posteriors for all parameters, enabling accurate means, uncertainties, and correlations to be recovered directly from imaging and photometry. For emission line fluxes (H$\alpha$, H$\beta$, [N~\textsc{ii}], [O~\textsc{iii}]), we follow the same two-stage setup as the first branch. We first train an MLP that predicts the mean fluxes in log1p space and then model the residuals using a 4D conditional RealNVP flow (12 affine coupling layers) conditioned on the encoder representation. This captures the joint posterior distribution over all four emission-line fluxes.

To train the VAE part of the model, we use an A100 GPU on google colab the required a training time of 1.5 hours. For the NF part, we trained on a T4 GPU that required approximately 30 minutes. 

\section{Results}

\begin{table}[h!]
\centering
\caption{\(R^2\) scores for predicting galaxy redshift, stellar mass, SFR, BH mass, and metallicity (12 + log(O/H)) across different studies in comparison to ours.}
\vspace{3pt}
\begin{tabular}{lccccc}
\hline\hline
\textbf{Method} & \(z\) & Mass & SFR & BH Mass & Metallicity \\
\hline
(r,g,z) Photometry + MLP [\cite{parker2024astroclip}]              & 0.68 & 0.67 & 0.34 & N/A & 0.41 \\
Image Embedding + MLP [\cite{parker2024astroclip}]                  & 0.78 & 0.73 & 0.42 & N/A & 0.43 \\
Image Embedding + kNN [\cite{parker2024astroclip}]                  & 0.79 & 0.74 & 0.44 & N/A & 0.44 \\
Image Embedding [\cite{gagliano2023physics}]                       & 0.83 & 0.75 & N/A  & N/A & N/A  \\
\textbf{Image Embedding + Photometry + NF (ours)} & \textbf{0.80} & \textbf{0.85} & \textbf{0.76} & \textbf{0.67} & \textbf{0.76} \\
\textbf{Photometry + NF (ours)}            & \textbf{0.72} & \textbf{0.80} & \textbf{0.75} & \textbf{0.62} & \textbf{0.65} \\
\hline
\label{tab:R^2}
\end{tabular}
\end{table}

Figure~\ref{fig:vae-dnn-architecture} shows predicted versus true values for $M_\star$,SFR, $z$, $M_{\mathrm{BH}}$, and metallicity. Our model achieves strong performance, with $R^2 = 0.85$ for stellar mass, $0.76$ for SFR, $0.80$ for redshift, $0.67$ for black hole mass, and $0.76$ for metallicity. The points cluster closely around the one-to-one line, showing that the model reliably predicts each parameter across the observed range of values.

\begin{figure}[t]
  \centering
  \begin{subfigure}{\linewidth}
    \centering
    \includegraphics[width=.99\linewidth,
  trim={0 0 0 0}, clip]{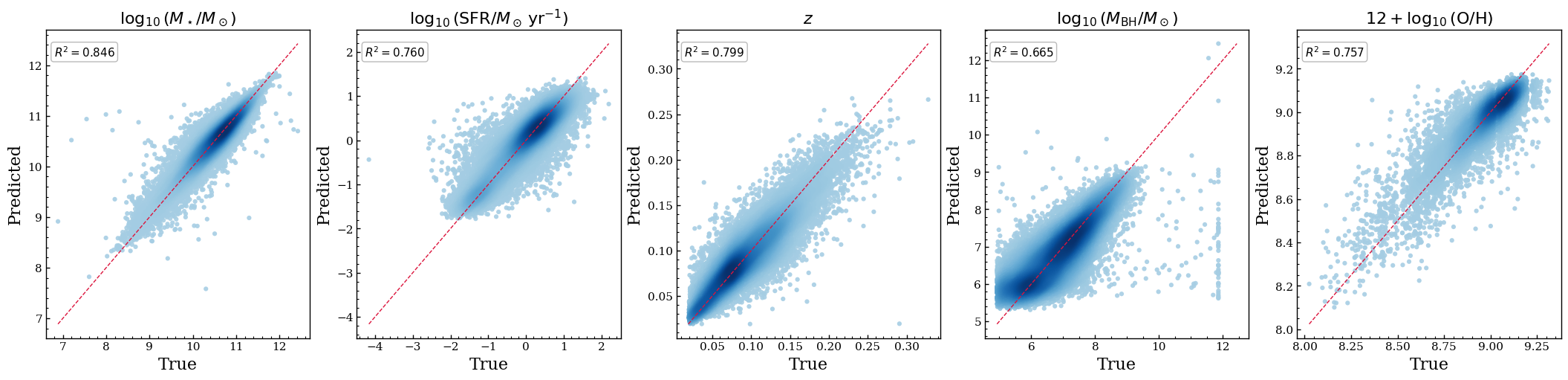} 
    \caption{Stellar mass ($\log M_\ast$), SFR, redshift ($z$), BH mass ($\log M_{\rm BH}$), and metallicity ($12+\log\mathrm{(O/H)}$).}
    \label{fig:pred_true_params}
  \end{subfigure}

  \vspace{3pt} 

  \begin{subfigure}{\linewidth}
    \centering
    \includegraphics[width=.99\linewidth]{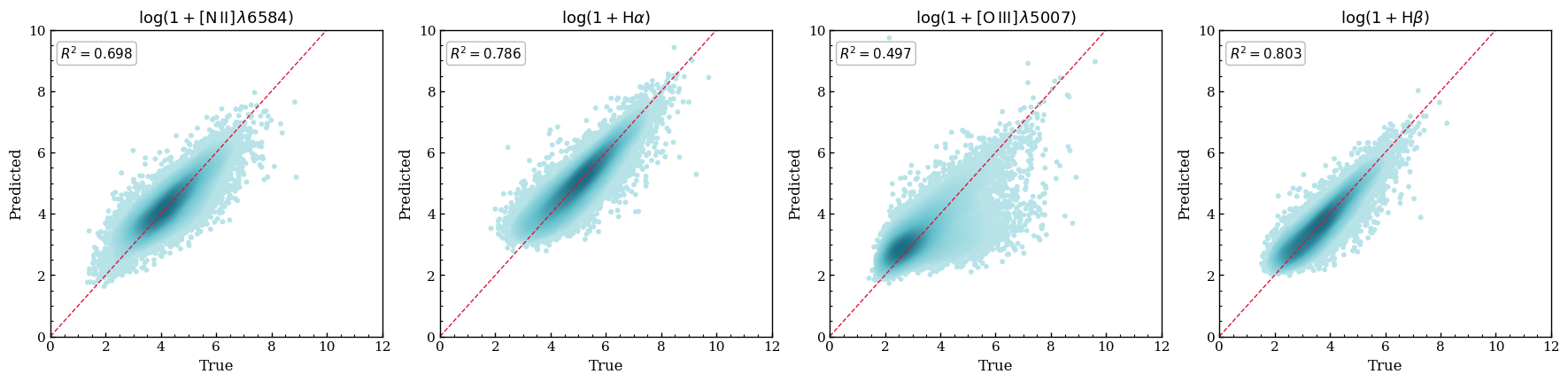} 
    \caption{Emission-line fluxes in log space: [N\,\textsc{ii}]$\lambda6584$, H$\alpha$, [O\,\textsc{iii}]$\lambda5007$, H$\beta$ (units $10^{-17}\,\mathrm{erg\,s^{-1}\,cm^{-2}}$).}
    \label{fig:pred_true_lines}
  \end{subfigure}

  \caption{Predicted versus true values for galaxy properties and emission-line fluxes. Each panel compares model predictions against ground truth, with the dashed red line indicating the one-to-one relation. }
  \label{fig:pred_vs_true_all}
  
\end{figure}

Predictions for nebular emission lines are shown in Figure~\ref{fig:pred_true_lines}. Balmer lines are accurately reproduced ($R^2 = 0.79$--$0.80$), while [N\,\textsc{ii}]$\lambda6584$ achieves moderate accuracy ($R^2 = 0.70$) and [O\,\textsc{iii}]$\lambda5007$ remains more challenging ($R^2 = 0.50$), reflecting its stronger dependence on ionization conditions not fully captured by global properties.

Table~\ref{tab:R^2} compares our results with previous studies. Our model outperforms both MLP and kNN baselines across all parameters, particularly for SFR ($R^2 = 0.76$ vs.\ 0.34--0.44 in prior work). This improvement stems from the NF component, which captures the underlying degeneracies among stellar mass, redshift, and SFR that deterministic regressors often fail to model. Even when trained on photometric inputs alone, the NF framework achieves higher accuracy for SFR and metallicity than previous image-embedding methods, highlighting the advantages of probabilistic modeling. We also tested the accuracy of our binary classifier using a confusion matrix, which showed that the model is approximately 84\% accurate in distinguishing galaxies with and without an observed metallicity measurement. The $R^2$ plot for black hole mass suggests that the model predicts lower values in cases where the catalog reports extremely large masses ($10^{11}$–$10^{12},M_\odot$), which are unlikely to be physical, indicating that the model may be capturing more realistic behavior than the underlying data. To assess uncertainty, we decompose the total predictive spread into aleatoric and epistemic components (Table~\ref{tab:uncertainty_two_rows}). The results show that aleatoric uncertainty, which captures intrinsic data scatter, dominates across all parameters, while the epistemic component remains small. Metallicity and redshift remain the most tightly constrained, while black-hole mass and [O III] 5007 flux show larger total uncertainties.


\captionsetup[subfigure]{justification=centering}

\begin{table*}
\centering
\small
\caption{Aleatoric and epistemic uncertainty (validation set). Emission-line values are reported in log1p space.}
\setlength{\tabcolsep}{3.5pt} 
\resizebox{\textwidth}{!}{%
\begin{tabular}{lccccccccc}
\hline
& $M_{\rm BH}$ [$M_\odot$] & $\log M_\star$ & $12+\log(\mathrm{O/H})$ & $\log \mathrm{SFR}$ & $z$ & H$\alpha$ & H$\beta$ & [N\,\textsc{ii}]\,6584 & [O\,\textsc{iii}]\,5007 \\
\hline
$\sigma_{\rm ale}$  & 0.589 & 0.191 & 0.134 & 0.327 & 0.018 & 0.427 & 0.381 & 0.427 & 0.611 \\
$\sigma_{\rm epi}$  & 0.034 & 0.012 & 0.010 & 0.019 & 0.001 & 0.027 & 0.026 & 0.027 & 0.045 \\
\hline
\end{tabular}%
}
\label{tab:uncertainty_two_rows}
\end{table*}

\vspace{0.5cm}
\textbf{Latent Space Interpretability}

\begin{figure}[htbp]
  \centering

  \begin{subfigure}{0.49\linewidth}
    \centering
    \includegraphics[width=.99\linewidth]{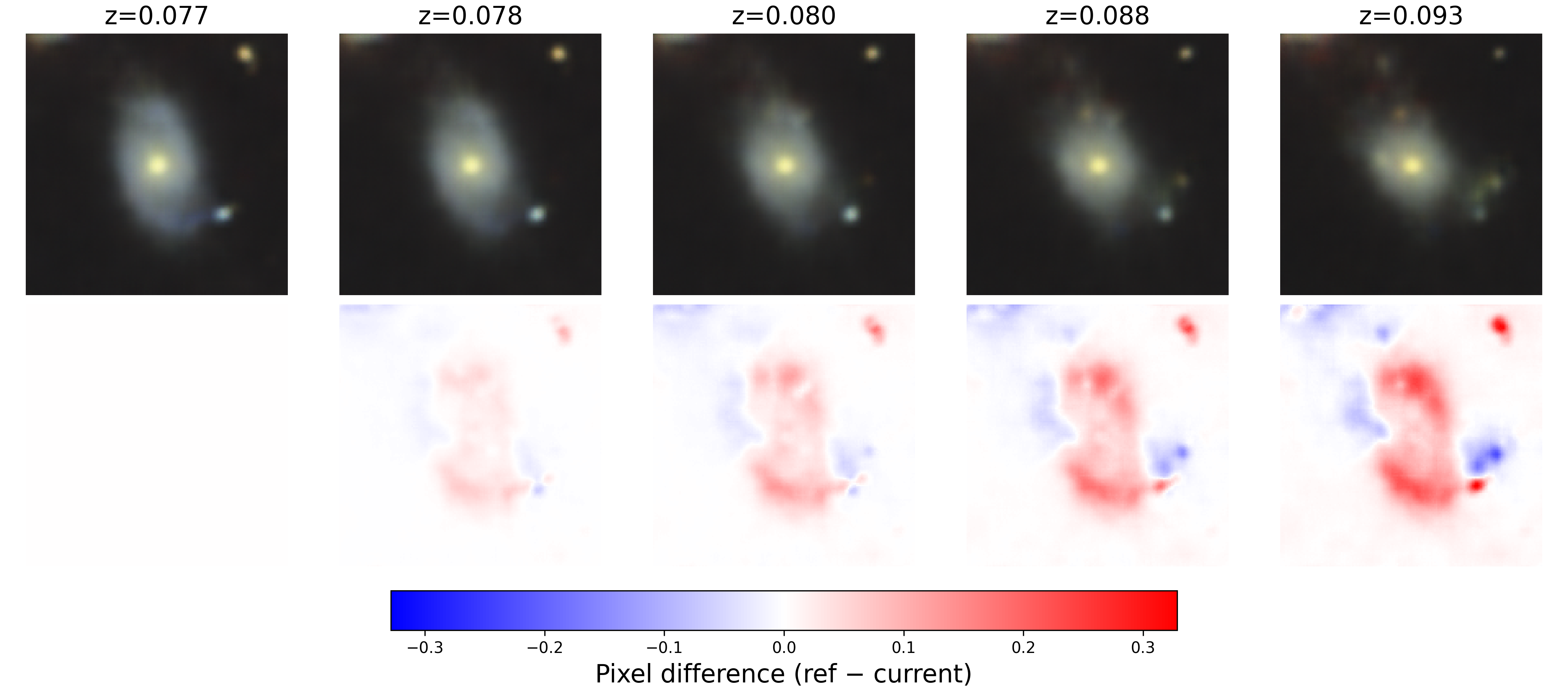}
    \caption{Redshift}
    \label{fig:latent:z}
  \end{subfigure}\hfill
  \begin{subfigure}{0.49\linewidth}
    \centering
    \includegraphics[width=.99\linewidth]{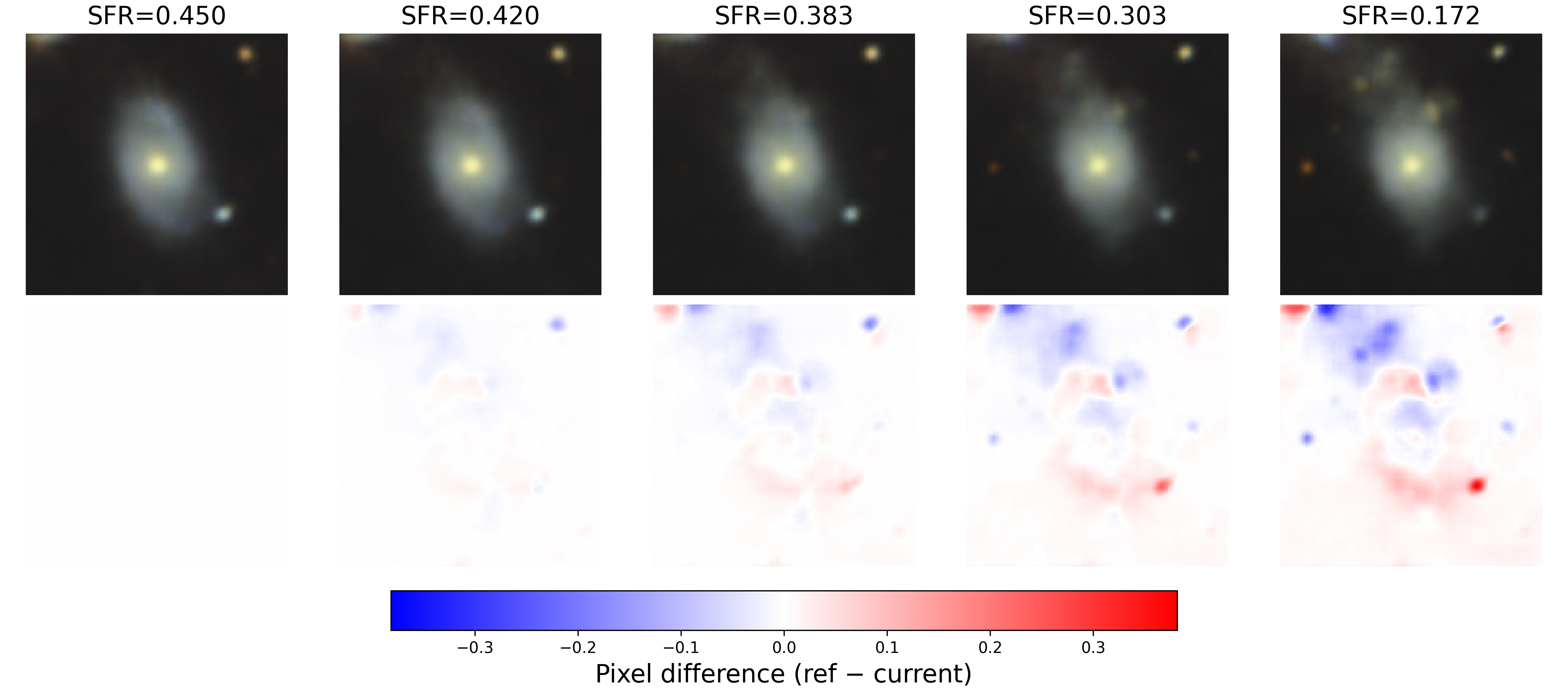}
    \caption{$\mathrm{log SFR(M_{\odot}/yr)}$}
    \label{fig:latent:SFR}
  \end{subfigure}

  \caption{Perturbational decoding of redshift and SFR of a representative galaxy. As we increase redshift, we see the galaxy getting smaller and farther away. As we decrease SFR, we see the galaxy obtaining a yellower bulge which is a common characteristic of galaxies with lower SFR.}
  \label{fig:latent}
\end{figure}

\begin{figure}[!htbp]
 
  \centering
  \begin{subfigure}[t]{0.23\textwidth}
    \includegraphics[width=\linewidth]{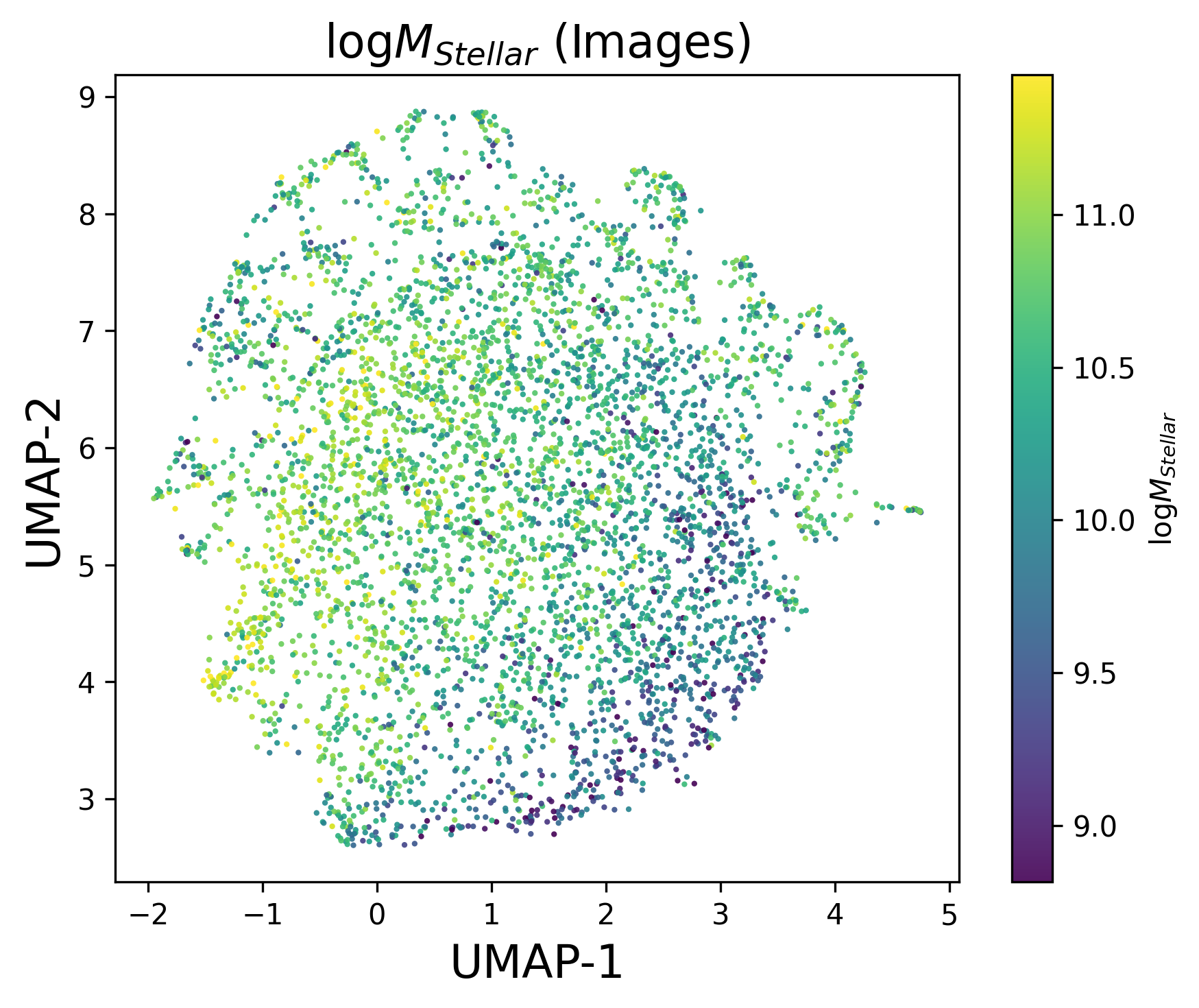}
    \caption{Stellar mass (images)}
    \label{fig:umap-img-mass}
  \end{subfigure}\hspace{0.008\textwidth}
  \begin{subfigure}[t]{0.23\textwidth}
    \includegraphics[width=\linewidth]{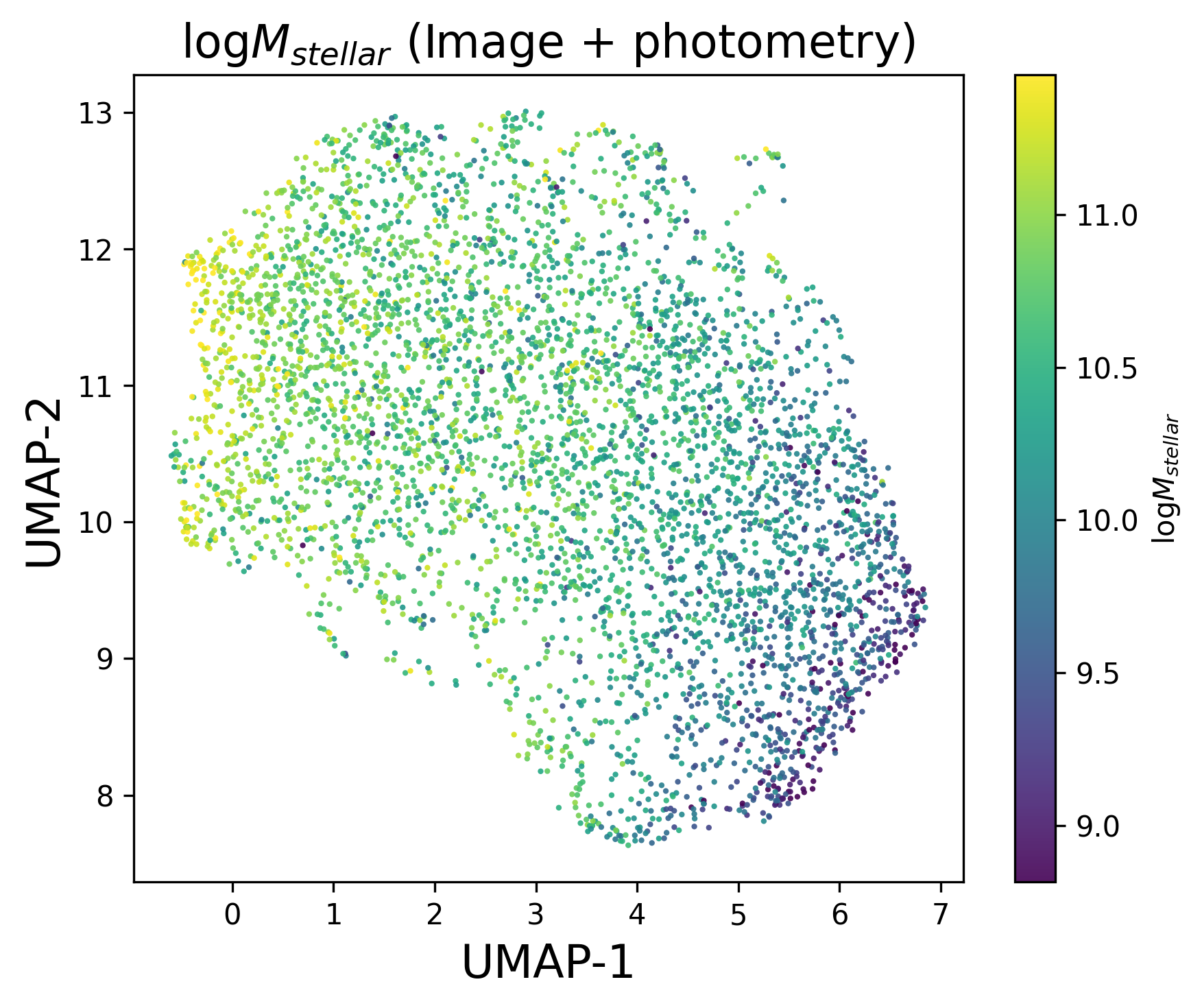}
    \caption{Stellar mass (image+phot)}
    \label{fig:umap-imgphot-mass}
  \end{subfigure}
  \hspace{0.035\textwidth}
  \begin{subfigure}[t]{0.23\textwidth}
    \includegraphics[width=\linewidth]{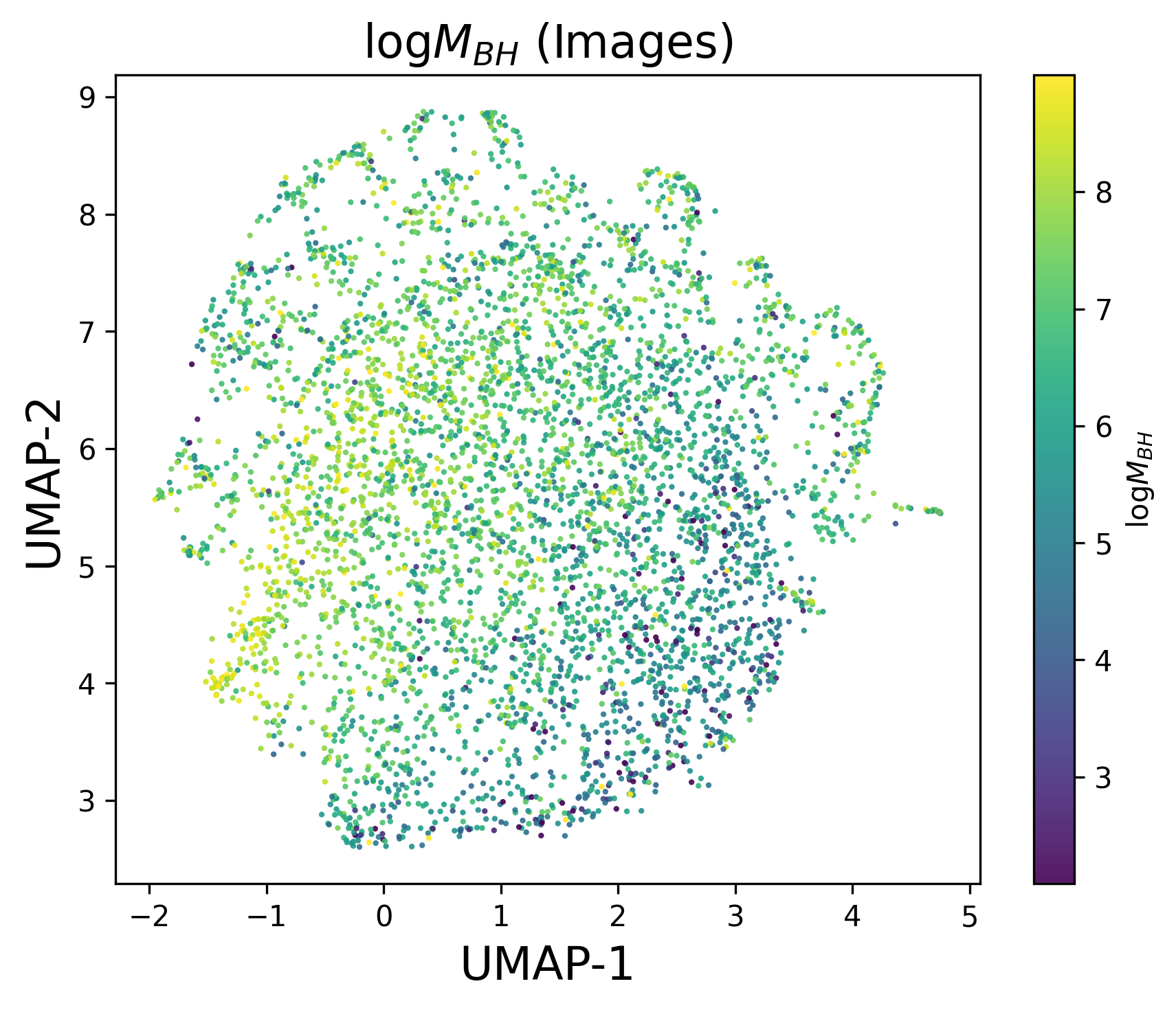} 
    \caption{BH mass (images)}
    \label{fig:umap-img-bh}
  \end{subfigure}\hspace{0.008\textwidth}
  \begin{subfigure}[t]{0.23\textwidth}
    \includegraphics[width=\linewidth]{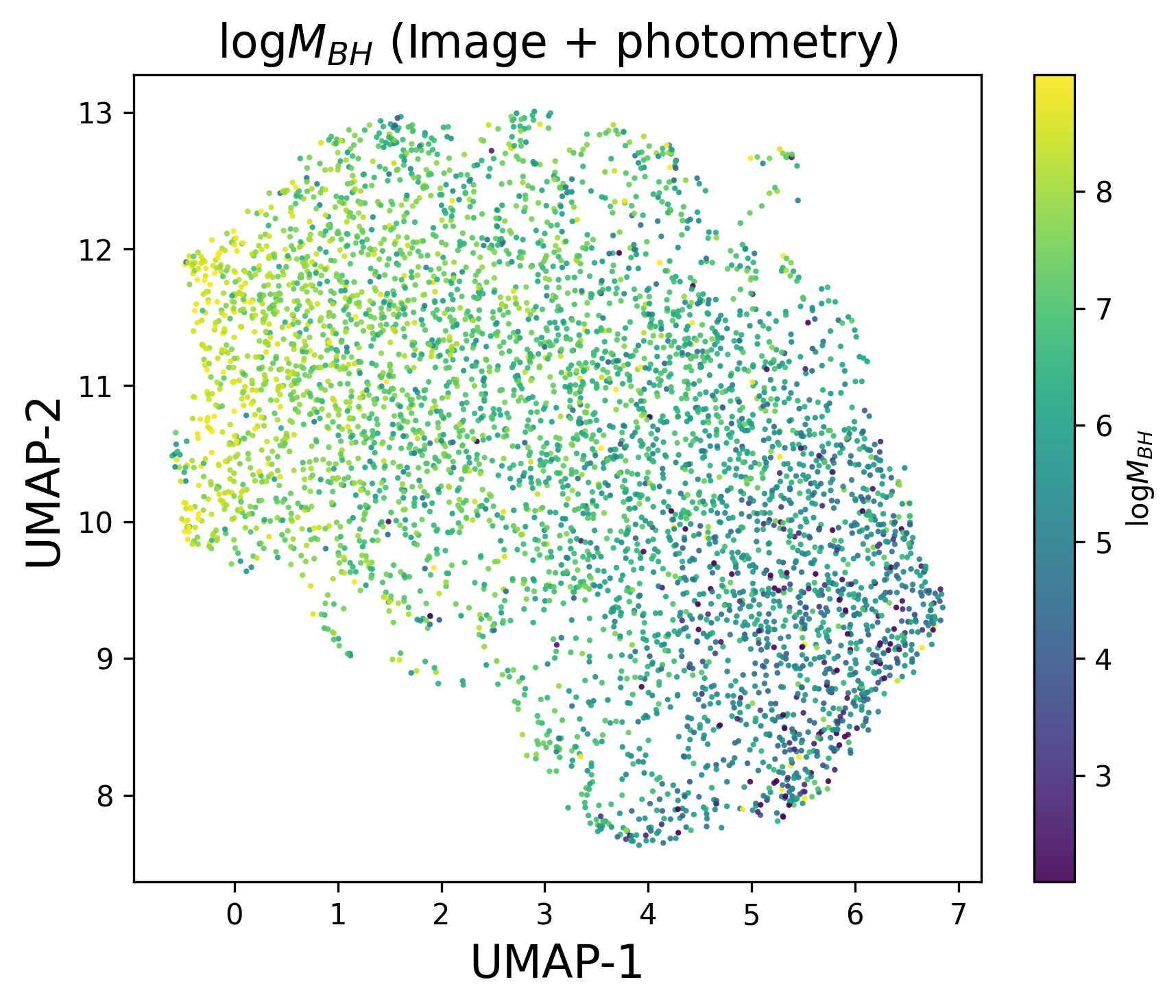}
    \caption{BH mass (image+phot)}
    \label{fig:umap-imgphot-bh}
  \end{subfigure}

  \caption{UMAP embeddings of galaxies when only images are used (a, c) and when images are combined with photometric information (b,d) for stellar mass and mass of central Black hole. For both cases, photometry significantly helps distinguish high mass and low mass cases in UMAP embeddings highlighting photometry's role in parameter inference.}
  \label{fig:umap}
\end{figure}

From the trained VAE, we obtain 32 latent dimensions. To identify which latent variables are most relevant for a given physical parameter, we perform a simple linear regression between the latent vectors and each predicted quantity. For each parameter, we select the three latent dimensions with the largest regression coefficients. All other latent dimensions are kept fixed, and we perturb the selected three within a small range while decoding through the VAE’s decoder to generate new synthetic galaxy images. This method aims to capture the morphological changes represented by changing each physical parameter. Figure~\ref{fig:latent} shows such an example for redshift and SFR. In Figure~\ref{fig:umap} we see UMAP \citep{mcinnes2018umap} embeddings for $M_{\star}$ and $M_{\mathrm{BH}}$. For Figure~\ref{fig:umap}(a,c) we use latent dimensions of images and for Figure~\ref{fig:umap}(b,d) we add photometric information of galaxies. Combining photometry enhances the separation of high and low mass galaxies. 

\section{Conclusion and Future Work}

We introduced a probabilistic framework based on Normalizing Flows that predicts global galaxy properties and emission-line fluxes directly from imaging data. In our tests, the method recovers stellar mass, star-formation rate, redshift, black hole mass, metallicity, and multiple nebular lines with good accuracy, and performs noticeably better than baseline MLP and kNN models. Because the approach learns a joint distribution rather than independent regressions, it captures correlations among physical quantities and produces uncertainties that reflect the data.

There remain several natural directions for future work. Our analysis so far use SDSS DR1, which is fairly shallow and noisy compared to later data releases. Moving to DR17 should provide higher-quality spectra and broader coverage, which will help with both training and validation. We also aim to push the model to a wider redshift range so that it can be used for earlier cosmic epochs and compared with upcoming survey data. Finally, although the VAE works well for encoding images, it can smooth out small-scale structure when the input is noisy. Diffusion-based models might be more effective at preserving these details, and we plan to explore them as a replacement for the VAE component.

\section{Acknowledgements}
    We thank our friends \& colleagues at AstroAI for their constructive feedback and support throughout the project. A.A.S. acknowledges support from Bryn Mawr College. S.S.M. acknowledges support from Colgate Research Council. 

\appendix

\section{Posterior Distribution Example}

\begin{figure}[htbp]
  \centering

  \begin{subfigure}{0.49\linewidth}
    \centering
    \includegraphics[width=.99\linewidth,
  trim={0 0 0 18pt}, clip]{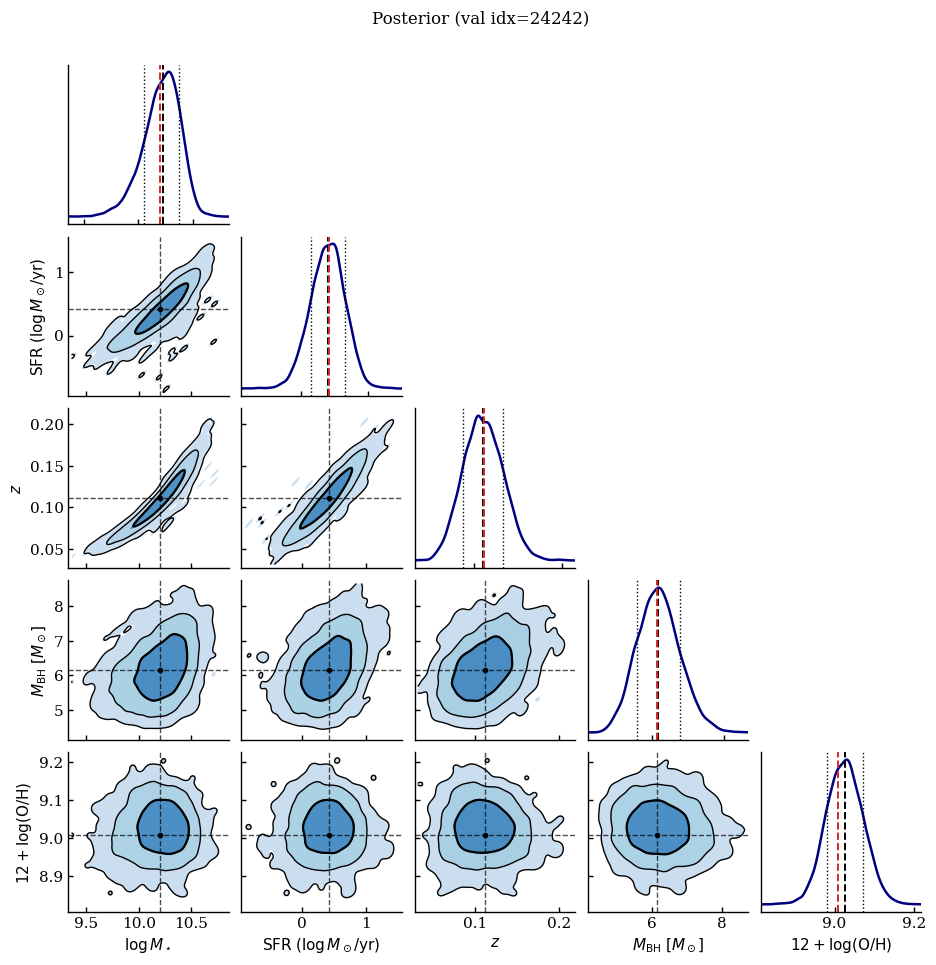}
    \caption{Physical parameters (stellar mass, SFR, $z$, BH Mass, Metallicity).}
    \label{fig:corner:params}
  \end{subfigure}\hfill
  \begin{subfigure}{0.49\linewidth}
    \centering
    \includegraphics[width=.99\linewidth,
  trim={0 0 0 18pt}, clip]{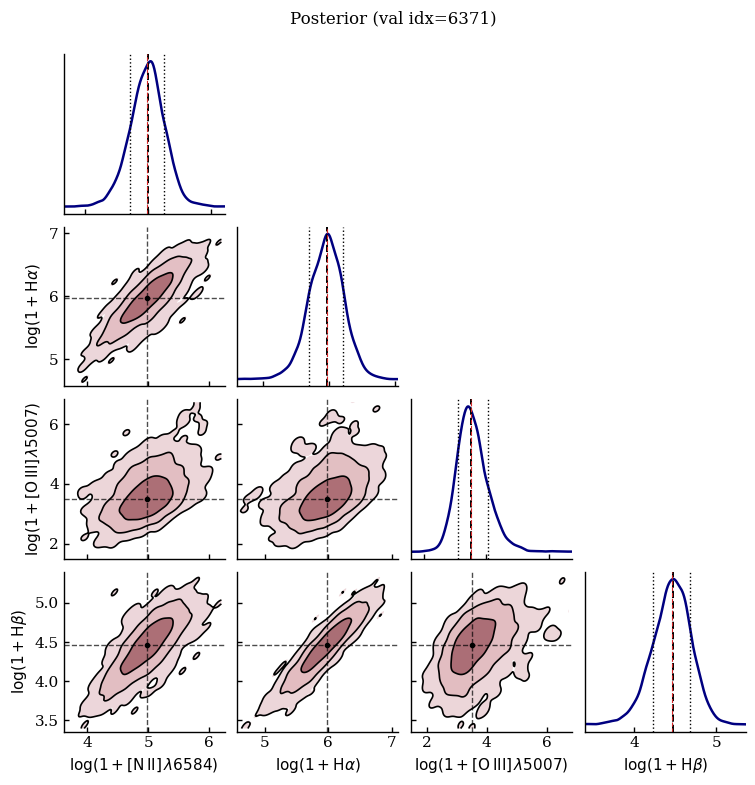}
    \caption{Emission-line fluxes: [N\,\textsc{ii}]$\lambda6584$, H$\alpha$, [O\,\textsc{iii}]$\lambda5007$, H$\beta$.}
    \label{fig:corner_lines}
  \end{subfigure}

  \caption{Posterior distributions for a representative galaxy. Diagonal panels show marginalized 1D posteriors; off-diagonals show 2D contours. 
  (a) Correlations between stellar mass, SFR, and redshift. 
  (b) Joint posterior over emission-line fluxes (log$(1+\mathrm{flux})$ units).}
  \label{fig:corner}
\end{figure}

Figure~\ref{fig:corner} shows joint posteriors for a representative galaxy. 
Diagonal panels give the marginalized 1D posteriors; the black bands mark the model’s \(1\sigma\) interval and the red dashed line marks the true value. 
The off-diagonal panels show the covariances between parameters. 
In panel (a), the \(\log M_\star\)–\(\log\mathrm{SFR}\) posterior is elongated along the star–forming “main sequence,” consistent with well–known trends and their redshift evolution \citep[e.g.,][]{noeske2007star,whitaker2012star}. 
Both \(M_\star\) and SFR co–vary with \(z\), reflecting color–magnitude degeneracies in photometric inference \citep[e.g.,][]{benitez2000bayesian,dahlen2013critical}. 
In panel (b), the Balmer lines (H\(\alpha\), H\(\beta\)) are tightly correlated while the [N\,\textsc{ii}]\,6584 and [O\,\textsc{iii}]\,5007 posteriors show structure expected from metallicity and ionization-parameter variations. \citep{baldwin1981classification}. 
Together, these corner plots illustrate that the flow recovers expected uncertainties and covariances among the parameters.

\bibliographystyle{unsrt}   
\bibliography{neurips_2024}

\end{document}